# Theory about Atlas morphology
# (Saturn moon)

Enrique Ordaz Romay[1]

*Facultad de Ciencias Físicas, Universidad Complutense de Madrid*

## Abstract

On June 12, 2007 the Cassini probe sent the images of a small moon of Saturn called Atlas which is located between the ring A and the small ring R/2004 S 1. These images have shown that the Atlas morphology is very different from other moons of similar dimensions. In the present article we propose a reasonable theory, to that we denominated "*flying dune*", that explains its morphologic characteristics from its magnitudes like mass, diameters and orbital radius, as well as its orbital position and the interpretation of the images caught by the Cassini probe.

---

[1] eorgazro@cofis.es

# Introduction.

In 1980, Voyager 1 probe flew over Saturn sending images of the planet, its rings and moons. Richard J. Terrile, in October, 1980 discovered in these images a new moon that orbited a little beyond to the ring A, to that called 1980 S28 provisionally [1]. Shortly after, the definitive name were Atlas [2] (see image 1).

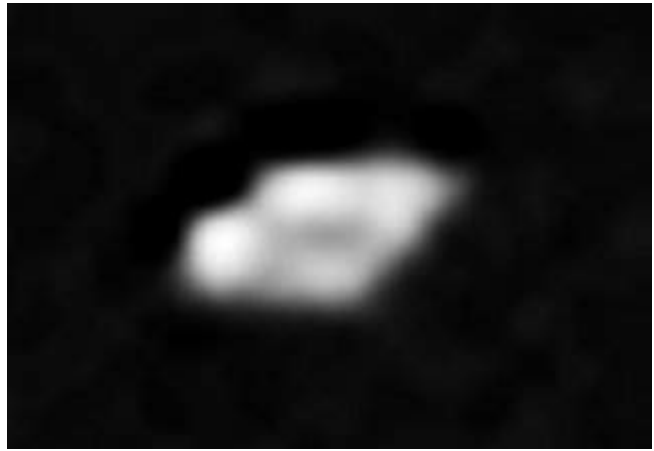

*Image 1. Photography obtained by the spaceship Voyager 1 on November 12, 1980. (Credit: NASA, JPL, SSI)*

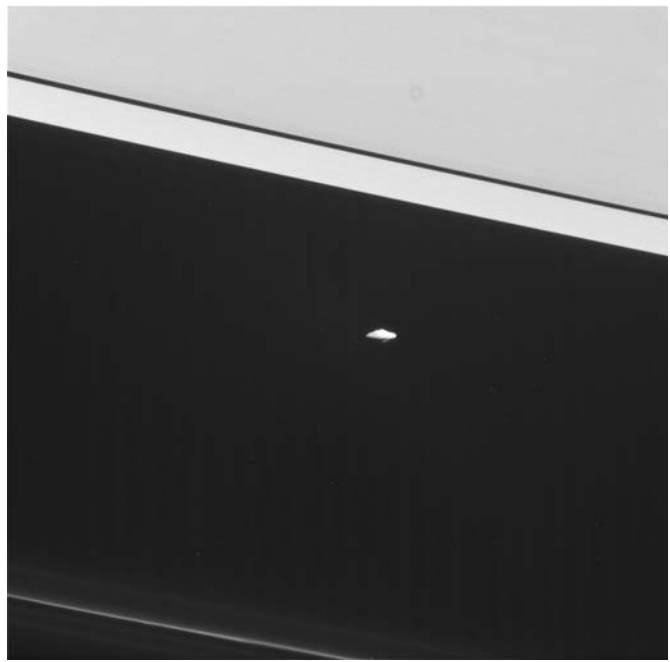

*Image 2. Photography obtained by the spaceship Cassini on June 8, 2005. (Credit: NASA, JPL, SSI)*

On June 8, 2005 the Cassini probe approached Atlas to a distance of 428,551 km obtaining different images [3] that showed a satellite with a big symmetry in their polar axis (see image 2).

An enlargement of this photography from 2005 reveals that the morphology of the satellite has disc form (see image 3).

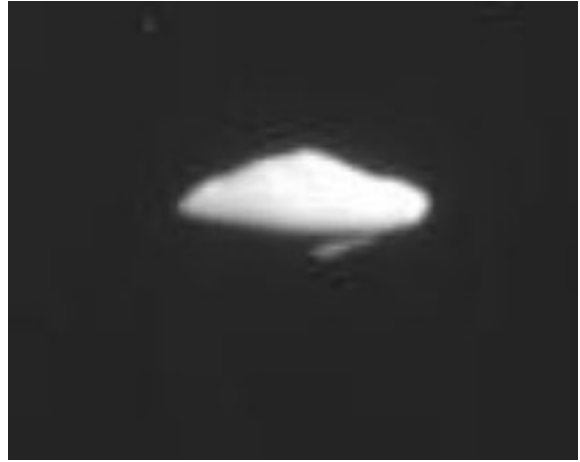

*Image 3. Atlas seen in an enlargement of the photography from June 8, 2005. (Credit: NASA, JPL, SSI)*

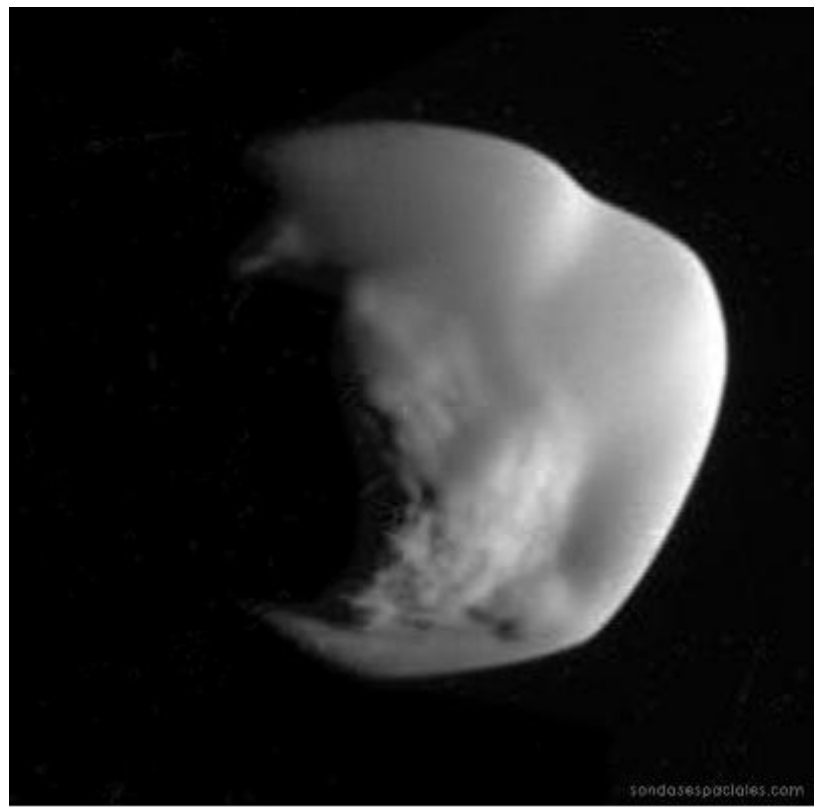

*Image 4. Atlas seen by Cassini on June 12, 2007. (Credit: NASA, JPL, SSI)*

On June 12, 2007 Cassini obtain an images series with better resolution form Atlas seen from the polar plane (see image 4).

# Morphologic differences between
# Atlas and other small moons.

If we compared the Atlas images from October 8, 2005 and June 12, 2007 with the images that we have from other "small" Saturn moons, we found several remarkable differences (see image 5)

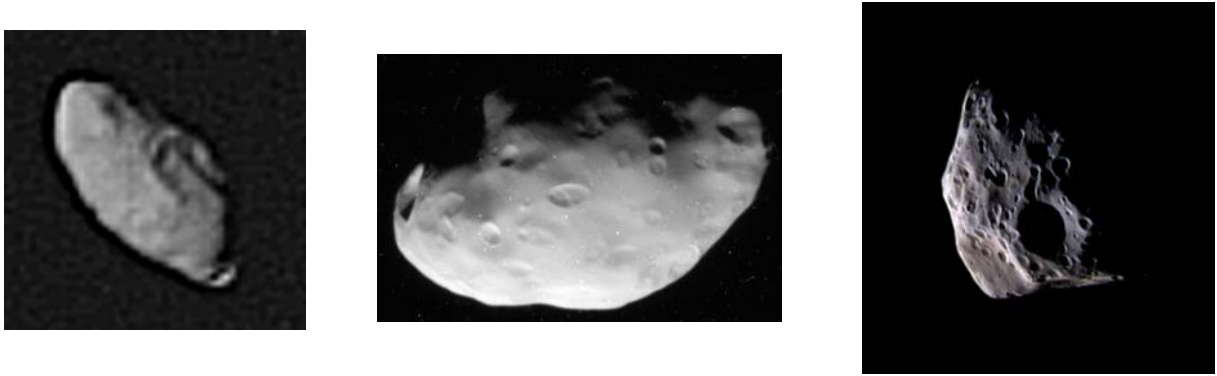

*Image 5. Prometheus, Pandora y Epimetheus. Moons of Saturn with near orbits and slightly greater than Atlas.*

The small satellites of other planets of the Solar System are very different too (see images 6 and 7).

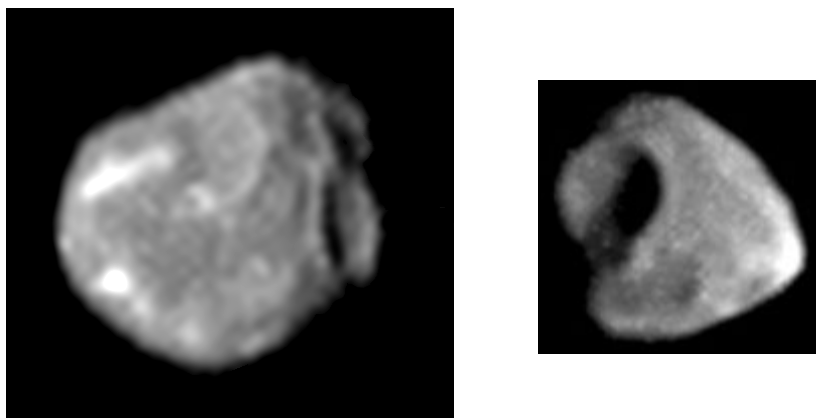

*Image 6. Amalhtea y Thebe. Moons of Jupiter.*

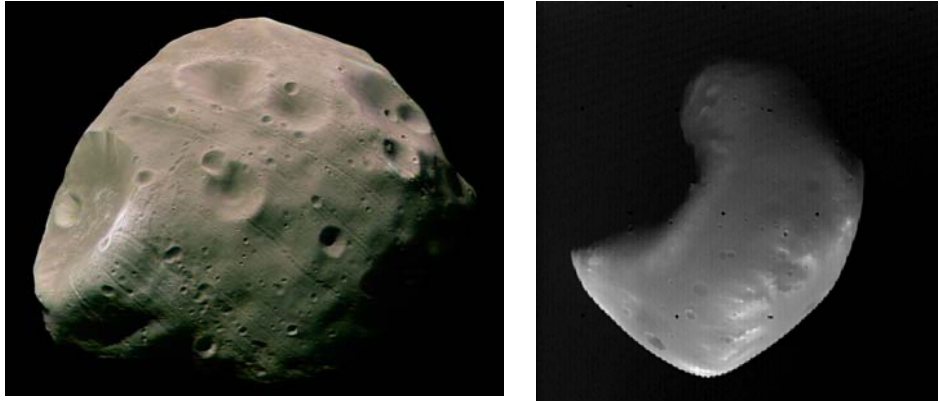
*Image 7. Phobos and Deimos. Moons of Mars.*

The differences between Atlas and other satellites of the Solar System are remarkable. Thus, Prometheus, Pandora and Epimetheus (Saturn satellites) with slightly greader and near orbits, or Amalthea and Thebe (Jupiter satellites) with very similar parameters, or Phobos and Deimos (Mars satellites) whose masses are minors than Atlas. We can observe the following differences:

- All the satellites are irregular whereas Atlas form has a central symmetry axe perpendicular to the plane of rotation and the Saturn ring.

- The other satellites show crater marks and Atlas does not have them..

- The surfaces of the other satellites seem rocks of rough aspect, whereas the Atlas surface seems polished.

## The Roche limit.

The "Roche limit" [4] is the distance within which, a satellite whose structure stays united only by its own gravity, begins to disintegrate itself, because the gravitational tidal forces from the planet to which it orbits and to the centripetal force of its rotation are greater than their gravitational force of cohesion.

Two different equations exist to calculate this limit depending the satellite is rigid or deformable. The difference between both equations only depends on a parameter that we will show by the letter $\delta$. So, the equation has the form:

$$d \approx \delta \cdot R \cdot \sqrt[3]{2\frac{\rho_M}{\rho_m}} \quad (1)$$

Siendo:

- $d$ = Roche limit
- $R$ = Radius of planet
- $\rho_M$ = Density of planet
- $\rho_m$ = Density of satellite.
- $\delta = \begin{cases} 1{,}260 & \text{for a solid satellite} \\ 2{,}423 & \text{for a fluid satellite} \end{cases}$

Our objective will be know, what is the value of the Roche limit for Atlas satellite if we considered that it is about a rigid satellite or fluid one. We will have to know the value the mass of the satellite for this.

## Problems about Atlas mass.

When we looked in the JPL.NASA web the Atlas characteristics we found that, the mass has diverse values depending on, which report from the web, we consult it:

- In "sse.jlp.nasa,gov" [2] the mass value is $8 \cdot 10^{17}$ kg

- In "saturn.jpl.nasa.gov"[3] the mass data is $2 \cdot 10^{15}$ kg

- On the other hand, in "nssdc.gsfc.nasa.gov"[5], although offering a value for the mass of $2 \cdot 10^{15}$ kg, it assigns a density of 500 $kg/m^3$, giving radial dimensions of 18.5 × 17.2 × 13.5 *km* (that we called *a*, *b* and *c* respectively). With these two data, the resulting mass for Atlas would be approximately of $9 \cdot 10^{15}$ *kg*.

- Finally, from the study made on 2006 by Spitale, J. N.; Jacobson, R. A.; Porco, C. C.; Owen, W. M., Jr. [6] can be deduced that $G \cdot m_{Atlas} = (0.44 \pm 0.04) \times 10^{-3}$ $km^3$ $s^{-2}$. As the universal gravitation constant is $G = (6.6742 \pm 0,0010) \cdot 10^{-11}$ $m^3$ $kg^{-1}$ $s^{-2}$ the result of the mass is: $(6,59 \pm 0,66) \cdot 10^{15}$ $kg$.

In summary, we have diverse values according to the source that we consult, whose values oscillate between $2 \cdot 10^{15}$ $kg$ and $8 \cdot 10^{17}$ $kg$ [7]

Nevertheless, this data that, at first, could seem that it has a secondary importance, in this case is fundamental for understand the images obtained by Cassini on June 12, 2007 [8].

On the contrary the values of the Atlas semiaxes could have been calculated with high accuracy. Atlas can be similar an ellipsoid of semiaxes *a*, *b* and *c* whose values are 18,500, 17,200 and 13,500 *km* respectively [9].

## The Roche limit for Atlas moon.

The value of Atlas mass is fundamental to know the Roche limit, if we considered it is rigid object or it is a fluid one.

The most trustworthy values for the mass are:

- $m_{Atlas} = 2 \cdot 10^{15}$ $kg$ according to the data from jpl.nasa.

- $m_{Atlas} = 6,6 \cdot 10^{15}$ $kg$ according to the calculations made by Spitale.

- $m_{Atlas} = 9 \cdot 10^{15}$ $kg$ according to the data of density from nssdc.gsfc.nasa.gov

With them we can make a table that relates the value of Roche Limit for a solid model as for the fluid model, for covers the mass values from this interval.

In order to apply the equation (1) we needed the density values of Saturn and Atlas: $\rho_M$ will be Saturn density which is calculated know Saturn is a revolution spheroid which equatorial and polar semiaxes are 120,536 and 108,728 *km* respectively. With these datas and knowing the Saturn mass is $5.688 \cdot 10^{26}$ *kg*, the density :

$$\rho_{Saturn} = \frac{m_{Saturn}}{\frac{4}{3}\pi \cdot r_{polar} \cdot r_{equatorial}^2} = 687,68 \; kg/m^3$$

On the other hand, $\rho_m$ will be the density of Atlas which corresponds to an ellipsoid of semiaxes 18.5 × 17.2 × 13.5 *km* (that we called *a*, *b* and *c* respectively) and we consider the mass of Atlas like variable. The result is:

$$\rho_{Atlas} = \frac{m_{Atlas}}{\frac{4}{3}\pi \cdot a \cdot b \cdot c} = 5,557 \cdot 10^{-14} \cdot m_{Atlas} \; kg/m^3$$

Finally the average Saturn radius calculates like the geometric average of the semiaxes[2] in the three space coordinates. Being Saturn a spheroid, its equation is:

$$r_{medio} = \sqrt[3]{r_{polar} \cdot r_{ecuatorial}^2} = 58.231,99 \; km$$

Replacing these values in the equation (1) we have left:

$$d \approx \delta \frac{1,34693 \cdot 10^{+13}}{\sqrt[3]{m_{Atlas}}} \qquad (2)$$

With these considerations we make a table with three columns, one for the values of the Atlas masses includes in interval $2 \cdot 10^{15} \; kg \leq m_{Atlas} \leq 9 \cdot 10^{15} \; kg$ and the others two columns for the two values of the parameter $\delta$.

---

[2] We used the geometric average because our objective is to calculate the sphere whose volume is equal to Saturn, being Saturn a spheroid.

| Atlas mass (kg) | Roche L. Solid ($\delta$=1,26) (km) | Roche L. fluid ($\delta$=2,423) (km) |
|---|---|---|
| $2.00 \cdot 10^{+15}$ | 134,701.44 | 259,033.01 |
| $2.50 \cdot 10^{+15}$ | 125,045.74 | 240,464.94 |
| $3.00 \cdot 10^{+15}$ | 117,672.55 | 226,286.18 |
| $3.50 \cdot 10^{+15}$ | 111,778.83 | 214,952.46 |
| $4.00 \cdot 10^{+15}$ | 106,912.60 | 205,594.63 |
| $4.50 \cdot 10^{+15}$ | 102,796,44 | 197,679.18 |
| $5.00 \cdot 10^{+15}$ | 99,248.87 | 190,857.15 |
| $5.50 \cdot 10^{+15}$ | 96,145.29 | 184,888.92 |
| $6.00 \cdot 10^{+15}$ | 93,396.76 | 179,603.46 |
| $6.50 \cdot 10^{+15}$ | 90,937.80 | 174,874.84 |
| $7.00 \cdot 10^{+15}$ | 88,718.92 | 170,607.88 |
| $7.50 \cdot 10^{+15}$ | 86,701.87 | 166,729.08 |
| $8.00 \cdot 10^{+15}$ | 84,856.59 | 163,180.57 |
| $8.50 \cdot 10^{+15}$ | 83,159.00 | 159,916.08 |
| $9.00 \cdot 10^{+15}$ | 81,589.59 | 156,898.07 |

*Table 1. Values of Roche Limit for Atlas like solid or fluid object, in function of its mass.*

As, the orbital Atlas radius is 137,670 km, it is easy calculate the exact value of the mass, over which, the satellite is outside the Roche Limit as much in the solid case as in the fluid case. Replacing in equation (2) *d* by the orbital radius and clearing the mass $m_{Atlas}$ is obtained:

$$m_{Atlas}(\text{rigid}; \delta = 1.26) > 1.8734 \cdot 10^{15} \text{ kg}$$

On the contrary, if we supposed that Atlas behaves as a fluid body the limit would calculate equal but applying to the constant $\delta = 2,423$. The result would be:

$$m_{Atlas}(\text{fluid}; \delta = 2.423) > 1.3322 \cdot 10^{16} \text{ kg}$$

It is interesting observe that, on the one hand, the propose mass by the JPL.NASA for Atlas only would be possible if Atlas were a solid body with forces of cohesion stronger than their gravitational forces.

On the other hand, Atlas cannot be taken as a totally deformable satellite because its Roche Limit for a fluid body is located very over the values that we are handling.

## Design about Atlas nature.

With the collected data, Atlas cannot be a deformable satellite, but seen its images (without craters and with symmetry in the rotation axis) it cannot either be a rigid satellite like Pandora, Thebe or Phobos.

Consequently, the only option that makes the observations and the calculations of the Roche Limits compatible is that Atlas is in an object with one part solid and in the other part deformable like a covered rock nucleus of a dust layer. If the nucleus are rigid, the deformation of the satellite is not complete like all of it was deformable.

The equations of the Roche Limit as they are expressed in the equation (1) have the form:

$$d \approx \delta \cdot R \cdot \sqrt[3]{2 \frac{\rho_M}{\rho_m}}$$

In which $\delta$ takes the values 1,26 ó 2,423.

Nevertheless, since the gravitational fields are additives, if we raised Atlas like an object formed by a rigid proportion and another deformable one, for their analysis we can factorize it in two parts. The Roche limit in this case can be considered, in first approach, of the form:

$$d \approx \left(1{,}26 \cdot p_{Atlas}(\text{rigid}) + 2{,}423(1 - p_{Atlas}(\text{rigid}))\right) \frac{1{,}3469 \cdot 10^{13}}{\sqrt[3]{m_{Atlas}}} \qquad (3)$$

Let $p_{Atlas}$(rigid) the satellite proportion of rigid mass (in fraction of one). That is, if satellite total mass ($m_{Atlas}$(total)) can be factorize in one part of solid mass ($m_{Atlas}$(rigid)) and other part of deformable mass ($m_{Atlas}$(deformable)) being fulfilled the relation:

$$m_{Atlas}(total) = m_{Atlas}(rigid) + m_{Atlas}(deformable) \text{ then } p_{Atlas}(rigid) = \frac{m_{Atlas}(rigid)}{m_{Atlas}(total)}$$

Let us suppose that Atlas has ellipsoid form because it is closely together of its Roche Limit ($d \approx 137{,}670$ km) according to his proportion of rigid and deformable mass. In this case we can calculate the proportion of the masses that compose it. Applying the equation (3) and clearing of the expression the proportion of rigid mass, it is obtained:

$$p_{Atlas}(rigid) = 2{,}066 - 8{,}7136 \cdot 10^{-6} \cdot \sqrt[3]{m_{Atlas}}$$

| Atlas Mass (kg) | Rigid mass proportion (%) |
|---|---|
| $2.00 \cdot 10^{+15}$ | 96.82 |
| $2.50 \cdot 10^{+15}$ | 88.34 |
| $3.00 \cdot 10^{+15}$ | 80.93 |
| $3.50 \cdot 10^{+15}$ | 74.30 |
| $4.00 \cdot 10^{+15}$ | 68.28 |
| $4.50 \cdot 10^{+15}$ | 62.74 |
| $5.00 \cdot 10^{+15}$ | 57.60 |
| $5.50 \cdot 10^{+15}$ | 52.79 |
| $6.00 \cdot 10^{+15}$ | 48.26 |
| $6.50 \cdot 10^{+15}$ | 43.98 |
| $7.00 \cdot 10^{+15}$ | 39.92 |
| $7.50 \cdot 10^{+15}$ | 36.04 |
| $8.00 \cdot 10^{+15}$ | 32.33 |
| $8.50 \cdot 10^{+15}$ | 28.77 |
| $9.00 \cdot 10^{+15}$ | 25.35 |

*Table 2. Relation between Atlas mass and its proportion of rigid mass for guarantee the gravitation stability.*

This relation allows us make a new table which indicates, if Atlas is near their Roche Limit, what must it be his proportion of rigid mass (see table 2).

Thus, for example, for a value of Atlas mass obtained in the study of Spitale and that corresponds to $6,6 \cdot 10^{15}$ kg the proportions between rigid and deformable mass are:

- $m_{rigid}$ > 43,15 %

- $m_{deformable}$ < 56,85 %

## Physical Atlas characteristics.

The made analysis of the images and calculations allow us to reach the conclusion that Atlas is an object formed by two types of material: one rigid and another deformable one. Since the condition of gravitational stability for an object thus would be the one of a central rock nucleus surrounded by a dust cloud in dune form. We will call to this model like "*flying dune*".

This model allows us to identify the rigid mass, calculated until now, with the mass of the rock nucleus, whereas the deformable mass corresponds to the mass of the dune. This is the notation that we will use from here (see figure 1).

In order to be able to analyze the gravitational Atlas field and if this model agrees with the images obtained by Cassini probe it is necessary to determine the density of the two parts of Atlas.

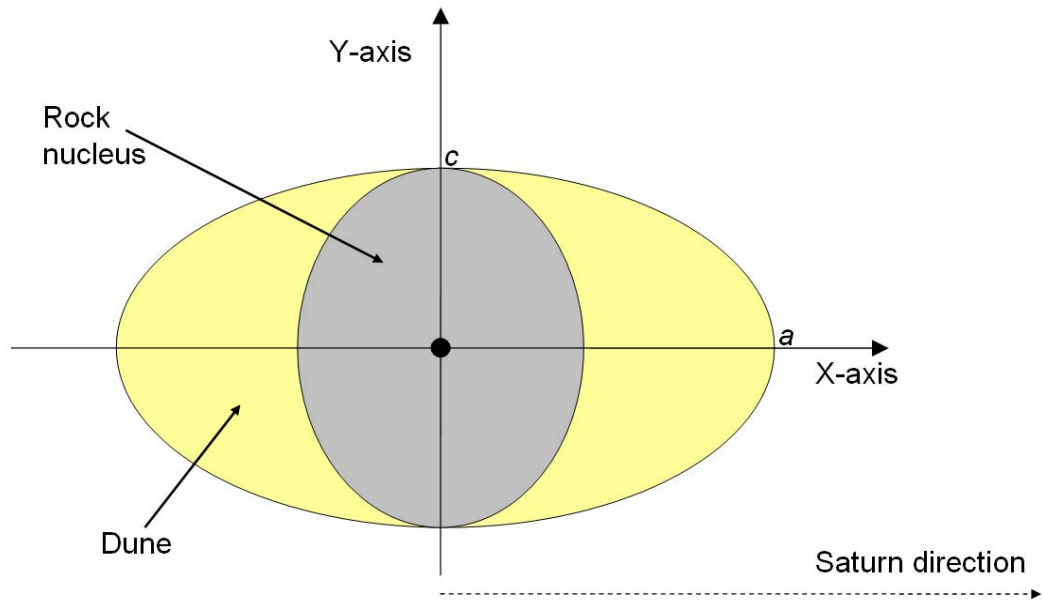

*Figure 1. Graphical model of the polar Atlas plane according to the model "volandora dune"*

We will take a hypothesis from reasonable, for this:

- Atlas mass is the one calculated on Spitae study and it is $6.6 \cdot 10^{15}$ kg,

- The Atlas images show, the satellite may situation closely of the Roche limit. For this reason and considering the previous mass we can determine that the mass of the rock ($m_{rock}$) corresponds to 43.15% of the total mass, whereas the mass of the dune ($m_{dune}$) would be 56.85%.

- The rock nucleus corresponds to a spheroid, one of whose semiaxes is $c = 13,500$ $m$ (the smaller Atlas semiaxis) and its other two semiaxes has similar values to each other. It is because in the Atlas images, in the poles, we can see dumpy texture of the rock. In addition, as we will see, the angular velocity of Atlas is displaced the dune towards the equator, clearing therefore the poles. That the equatorial poles of the rock are similar can suppose by the friction that polishes to the rock.

- The density of the rock nucleus is analogous to the one of any rocky satellite of similar dimensions. In our study we have considered an equal density to the Phobos satellite. That is to say, 1,900 $kg/m^3$

With these suppositions we can consider that the mass of the rock and the dune are:

- $m_{rock} = 0.4315 \cdot 6.6 \cdot 10^{15}\ kg = 2.848 \cdot 10^{15}\ kg$
- $m_{dunee} = 6.6 \cdot 10^{15}\ kg - 2.848 \cdot 10^{15}\ kg = 3.752 \cdot 10^{15}\ kg$

We have considered the density of the rock ($\rho_{rock}$) in 1,900 $kg/m^3$, in consequently its volume will be:

$$Volume_{rock} = \frac{m_{rock}}{\rho_{rock}} = 1.499 \cdot 10^{12}\ m^3$$

This volume must correspond to a spheroid with polar semiaxis equal to $c = 13,500\ m$ (polar Atlas semiaxis) and an equatorial radius ($r_{rock;eq}$) that we calculated:

$$\frac{4}{3}\pi \cdot 13,500 \cdot r_{rock;eq}^2 = 1.499 \cdot 10^{12} \rightarrow r_{rock;eq} = 5,148m$$

That is to say, with the suppositions that we have done, the rock nucleus has the next characteristics:

- Mass: $m_{rock} = 2,848 \cdot 10^{15}\ kg$
- Semiaxes = 13.500 × 5.148 × 5.148 $m$
- Density = 1.900 $kg/m^3$

With these data the volume of the dune will be:

$$Volume_{dune} = Volume_{Atlas} - Volume_{rock} = \frac{4}{3}\pi \cdot a \cdot b \cdot c - 1.499 \cdot 10^{12}\ m^3 = 1.649 \cdot 10^{13}\ m^3$$

To this volume corresponds a mass of $m_{dune} = 3.752 \cdot 10^{15}$ kg. Therefore the density of the dune ($\rho_{dune}$) is:

$$\rho_{dune} = \frac{m_{dune}}{Volume_{dune}} = 227.5 kg/m^3$$

Consequently the dust dune will have the next physical characteristics:
- Mass: $m_{dune} = 3.752 \cdot 10^{15}$ kg
- Volume: $Volume_{dune} = 1.649 \cdot 10^{13}$ m$^3$
- Density = 227.5 kg/m$^3$

## Gravitational Atlas field.

With these physical characteristics, the gravitational Atlas field, inside and outside the satellite consists of several equations that are described by sections. It is because the different parts from which the satellite consists have different densities, as well as each part has irregular form that we will consider like ellipsoids with the different semiaxes (see figure 2).

The general case would contemplate five zones from the center of masses to the space beyond the greater semiaxis. In the direction of the line that passes through the center of mass and Saturn, the intensity of gravitational field has the form that is seen in graph 1

In our model we are only interested in knowing the gravitational field on the surface and we restrict to the plane that passes through the poles and Saturn. In these conditions we have two zones according to the distance (*r*) to the center of mass:

- Dune II: $c \leq r \leq b$
- Dune III: $b \leq r \leq a$

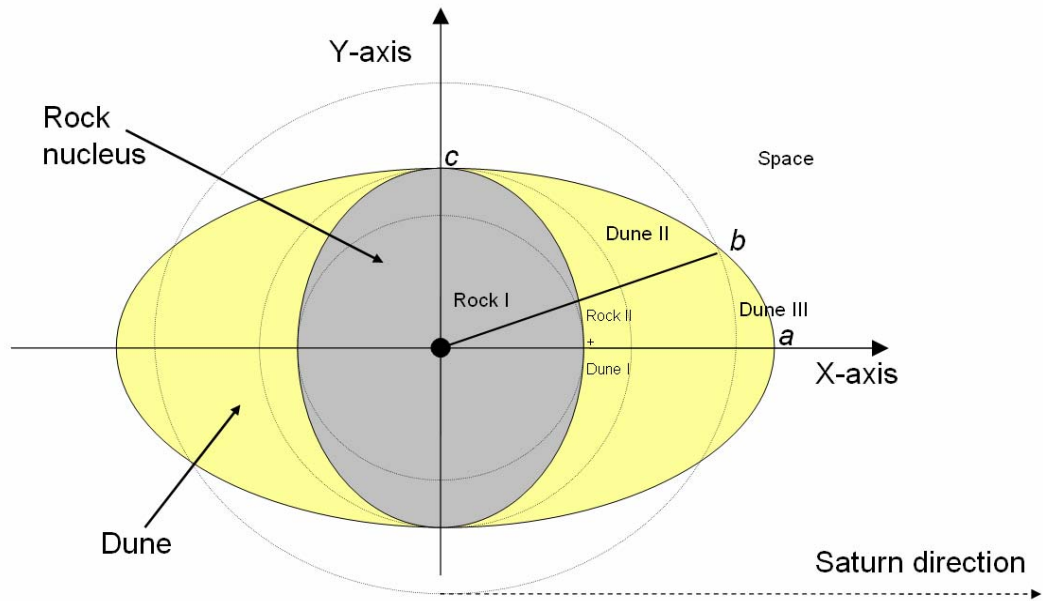

*Figure 2. Division of the zones of Atlas according to the model "flying dune"*

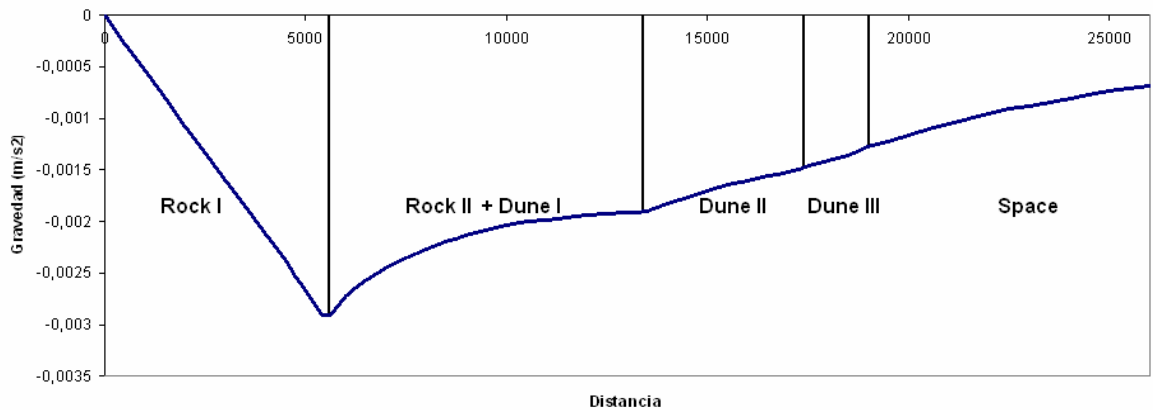

*Graph 1. Intensity of gravitational field in Atlas.*

The masses of each part of the dune, corresponding to the cut of the ellipsoid of semiaxes *a*, *b* and *c* with the sphere of radius *r* has a dependency of the form:

$$m_{II} = k_1 \cdot x^2 \quad \text{for} \quad c < r < b \quad \text{and} \quad m_{III} = k_2 \cdot x \quad \text{for} \quad b < r < a$$

Where we have considered that $k_1$ and $k_2$ are two slightly variant functions in $x$, that depend directly on the density of the dune and the semiaxes.

We must remember that we have made this approximation because at the end of the calculations in the surface equations we will have to readjust these values [3].

The gravitational intensity in both zones and on the surface we can estimate by the equations:

$$g_{DuneII} = -G\frac{m_{rock}}{r^2} - G\cdot\frac{k_1\cdot x^2}{r^2} + G\frac{k_1\cdot r^2_{rock;eq}}{r^2}$$

$$g_{DuneIII} = -G\frac{m_{rock}}{r^2} - G\cdot\frac{k_2\cdot x}{r^2} + G\frac{k_1\cdot r^2_{rock;eq}}{r^2}$$

The last term of the two previous expressions is added to counteract that in the volume of the rock nucleus we do not have any dune. We can write the previous expressions in more simplified form to defining the concept of effective mass of the nucleus ($m_n$) in the form:

$$m_n = m_{rock} - k_1\cdot r^2_{rock;eq}$$

Thus, the previous expressions take the form::

$$g_{DuneII} = -G\frac{m_n}{r^2} - G\cdot\frac{k_1\cdot x^2}{r^2}$$

$$g_{DuneIII} = -G\frac{m_n}{r^2} - G\cdot\frac{k_2\cdot x}{r^2}$$

---

[3] This approximation is based on a parallelepiped model. For revolution ellipsoid model we would have factors of form $(a^2 - x^2)^{3/2}$ that would complicate the model.

# Mathematical analysis of the "flying dune" model for Atlas.

**1.- Force analysis**

In Atlas model formed by a central rock and a encircle dust dune, the forces analysis in a point of the surface of the dune would be like in figure 3.

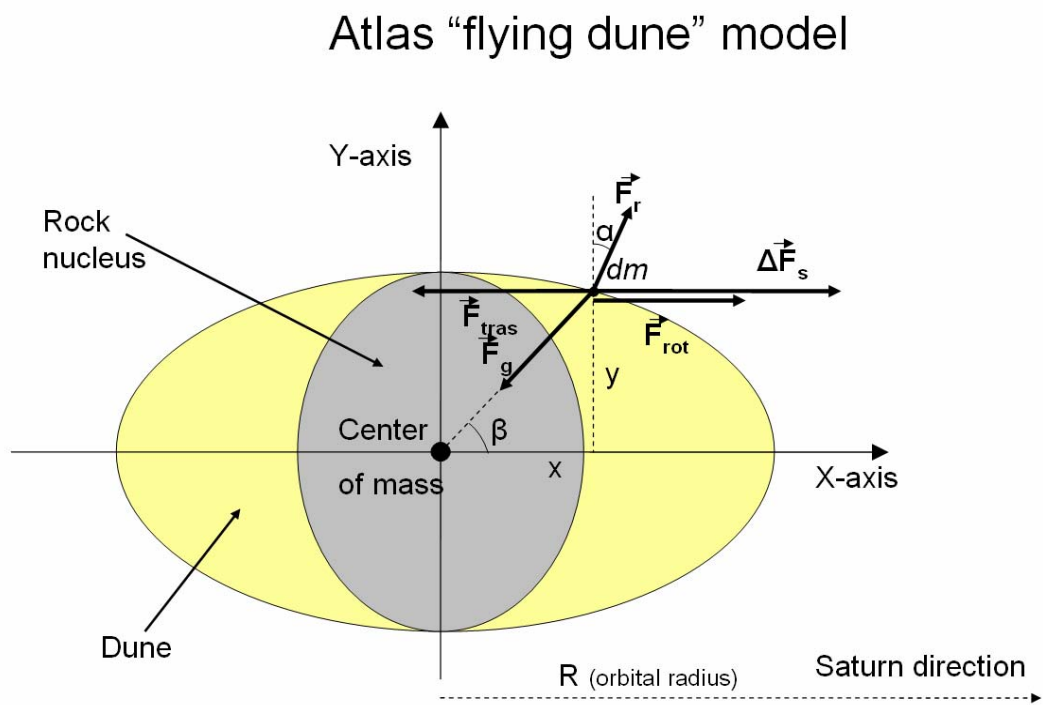

*Figure 3. Forces analysis in Atlas according to the model "flying dune"*

Where:

- $\vec{F}_g$ is the gravitation force of Atlas.
- $\vec{F}_{rot}$ is the centripetal force by Atlas rotation.
- $\vec{F}_{tras}$ is the centripetal force by Atlas translation.

- $\Delta \vec{F}_s$ is the difference of Saturn gravitational attraction between this point and the mass center. It is well-known as gravitational tidal force.
- $\vec{F}_r$ is the reaction of the surface caused by the pressure, friction, superficial tension of the dune (intermolecular forces, magnetic fields,…), etc.

- **The force of Atlas gravity** $\vec{F}_g$, as we saw in the previous section, applied on the surface, has two parts:

$$\vec{F}_{g;DuneII} = -G\left(\frac{m_n + k_1 \cdot x^2}{r^2}\right)\frac{dm}{r}(x,y)$$

$$\vec{F}_{g;DuneIII} = -G\left(\frac{m_n + k_2 \cdot x}{r^2}\right)\frac{dm}{r}(x,y)$$

- **The two centripetal forces:** one of rotation and another one of translation. Their equations would have to be:

$$\vec{F}_{rot} = dm \cdot w_1^2 \cdot (x,0)$$

$$\vec{F}_{tras} = -dm \cdot w_2^2 \cdot (R-x,0)$$

According to the data of the JPL.NASA the satellite Atlas spin is synchronous with its translation. It means that their angular velocities for rotation and translation are equal. That is to say, the relation is fulfilled $w_1 = w_2$ to which we will call simply $w$.

Replacing and adding it is: $\vec{F}_{tras} + \vec{F}_{rot} = -dm \cdot w^2 \cdot R \cdot (1,0)$

On the other hand, the eccentricity of the Atlas orbit is practically zero which indicates that in mass center the identity is verified between the gravitational intensity of Saturn and the centripetal acceleration:

$$w^2 \cdot R = G\frac{m_{Saturn}}{R^2} \rightarrow w = \sqrt{G\frac{m_{Saturn}}{R^3}}$$

Therefore: $\vec{F}_{tras} + \vec{F}_{rot} = -dm \cdot G \dfrac{m_{Saturn}}{R^2} \cdot (1,0)$

**- The tidal force** is analyzed starting from the idea that all the satellite is immersed in the gravitational Saturn field. This way, if the gravitational field were constant in all the satellite, any point of the satellite would be also seen affected by the gravity and in the mutual interactions between the parts of the satellite the Saturn gravity is cancelled.

Nevertheless, Atlas is sufficiently near Saturn like the differences of gravitational intensity that Saturn exerts between any point and the center mass are appreciable. This difference is that exerts gravitational force on the satellite to deform it.

The tidal force is then:

$$\Delta \vec{F}_s = G \dfrac{m_{Saturn}}{(R-x)^2} dm(1,0) - G \dfrac{m_{Saturn}}{R^2} dm(1,0)$$

**- The reaction force of the surface** is difficult to analyze in detail, because it implies many forces that altogether generate a tension superficial.

Nevertheless, on this force we only interest that it is perpendicular to the surface and therefore we can write it in the form:

$$\vec{F}_r = F_r (\sin \alpha, \cos \alpha)$$

The interest about this expression is because "tan $\alpha$" is the negative slope of the surface. If Atlas surface cut the plane of the semiaxes *a* and *c* in the equation curve *y* = *f(x)* then *y'* = – tan $\alpha$. This will help us to find the equation of the surface.

## Surface equations.

The balance state of the dune surface takes place when the sum of the forces is equal to zero. That is to say:

$$\text{Dune II} \begin{cases} x-axes: F_r \sin\alpha = G\left(\dfrac{m_n - k_1 \cdot x^2}{r^3}\right) \cdot x \cdot dm - G\dfrac{m_{Saturn}}{(R-x)^2} dm \\ y-axes: F_r \cos\alpha = G\left(\dfrac{m_n - k_1 \cdot x^2}{r^3}\right) \cdot y \cdot dm \end{cases}$$

$$\text{Dune III} \begin{cases} x-axes: F_r \sin\alpha = G\left(\dfrac{m_n - k_2 \cdot x}{r^3}\right) \cdot x \cdot dm - G\dfrac{m_{Saturn}}{(R-x)^2} dm \\ y-axes: F_r \cos\alpha = G\left(\dfrac{m_n - k_2 \cdot x}{r^3}\right) \cdot y \cdot dm \end{cases}$$

These are the parametric equations of a function $y = f(x)$ defined in the two zones. Dividing the two equations of each zone and remembering that $y' = -\tan\alpha$ we arrived at the two equations differentials:

$$\text{Dune II}: -y \cdot y' = x - \dfrac{m_{Saturn} \cdot r^3}{(R-x)^2 (m_n - k_1 \cdot x^2)} \tag{4}$$

$$\text{Dune III}: -y \cdot y' = x - \dfrac{m_{Saturn} \cdot r^3}{(R-x)^2 (m_n - k_2 \cdot x)}$$

These last equations can be integrated in the form:

$$-y \cdot y' = x - \dfrac{m_{Saturn} \cdot r^3}{(R-x)^2 h(x)} \rightarrow -y \cdot y' - x = -\dfrac{m_{Saturn} \cdot r^3}{(R-x)^2 h(x)} \rightarrow$$

$$-y \cdot dy - x \cdot dx = -\dfrac{m_{Saturn} \cdot r^3}{(R-x)^2 h(x)} \cdot dx \rightarrow \dfrac{1}{2} d(x^2 + y^2) = \dfrac{m_{Saturn} \cdot r^3}{(R-x)^2 h(x)} \cdot dx \rightarrow$$

$$\dfrac{1}{2} d(r^2) = \dfrac{m_{Saturn} \cdot r^3}{(R-x)^2 h(x)} \cdot dx \rightarrow \int \dfrac{dr}{r^2} = \int \dfrac{m_{Saturn}}{(R-x)^2 h(x)} \cdot dx \rightarrow -\dfrac{1}{r} + K = \int \dfrac{m_{Saturn}}{(R-x)^2 h(x)} \cdot dx$$

Where $K$ is the integration constant over $r$ variable. The $h(x)$ is:

$$Dune\ II: h_{DuneII}(x) = m_n - k_1 x^2$$
$$Dune\ III: h_{DuneIII}(x) = m_n - k_2 x$$

In order to find the solutions we must solve the integrals:

$$\int \frac{dx}{(R-x)^2 \cdot (m_n - k_1 \cdot x^2)} = \int \frac{A_1 x + B_1}{(R-x)^2} dx + \int \frac{C_1 x + D_1}{m_n - k_1 \cdot x^2} dx$$

$$\int \frac{dx}{(R-x)^2 \cdot (m_n - k_2 \cdot x)} = \int \frac{A_2 x + B_2}{(R-x)^2} dx + \int \frac{C_2}{m_n - k_2 \cdot x} dx$$

Where the constants are:

$$A_1 = \frac{-2}{k_1 R^3 - 3Rm_n} \quad B_1 = \frac{m_n - 3k_1 R^2}{k_1 R^2 (k_1 R^2 - 3m_n)} \quad C_1 = \frac{-2k_1}{k_1 R(k_1 R^2 - 3m_n)} \quad D_1 = \frac{m_n - R}{R^2 (k_1 R^2 - 3m_n)}$$

$$A_2 = \frac{k_2}{(k_2 R - m_n)^2} \quad B_2 = \frac{2k_2 R - m_n}{(k_2 R - m_n)^2} \quad C_2 = \frac{-k_2^2}{(k_2 R - m_n)^2}$$

Each integral separately is:

$$\int \frac{Ax + B}{(R-x)^2} dx = A\left(\frac{R}{R-x} + \ln(R-x)\right) + \frac{B}{R-x} + K$$

$$\int \frac{Cx + D}{m_n - k \cdot x^2} dx = \frac{C}{2} \ln(m_n - kx^2) + \frac{k \cdot D}{2} \sqrt{\frac{k}{m_n}} \cdot \ln\left(\frac{\sqrt{m_n} + x \cdot \sqrt{k}}{\sqrt{m_n} - x \cdot \sqrt{k}}\right) + K$$

$$\int \frac{C}{m_n - k \cdot x} dx = -\frac{C}{k} \ln(m_n - kx) + K$$

The result of replace all these expressions in the equations (4) leads to the exact solutions.

A form to simplify the solutions is remembering that $R \gg x$. In this case, returning to (4) we can do $(R - x) \approx R$ with which we have left:

$$-\frac{1}{r} + K \approx \frac{m_{Saturn}}{R^2} \int \frac{dx}{h(x)} = \begin{cases} DuneII : \frac{m_{Saturn}}{R^2} \int \frac{dx}{m_n - k_1 \cdot x^2} = \frac{m_{Saturn}}{2R^2 \cdot \sqrt{m_n}} \ln\left|\frac{\sqrt{m_n} - x\sqrt{k}}{\sqrt{m_n} + x\sqrt{k}}\right| \\ DuneIII : \frac{m_{Saturn}}{R^2} \int \frac{dx}{m_n - k_2 \cdot x} = -\frac{m_{Saturn}}{k_2 \cdot R^2} \ln\left|m_n - k_2 \cdot x\right| \end{cases}$$

Inasmuch as *Dune II* is defined between $0 < x < \sqrt{b^2 - c^2} = 10.657\ m$, we can approximate

$$\ln\left|\frac{\sqrt{m_n} - x\sqrt{k}}{\sqrt{m_n} + x\sqrt{k}}\right| \approx -\sqrt{\frac{k}{m_n}} \cdot x$$

In summary, the looked relations are:

$$DuneII : x^2 + y^2 = \frac{4R^4 \cdot m_n^2}{\left(m_{Saturn} \cdot x \cdot \sqrt{k_1} + K_1\right)^2}$$

$$DuneIII : x^2 + y^2 = \frac{k_2^2 \cdot R^4}{m_{Saturn}^2 \cdot \ln^2\left|m_n - k_2 \cdot x\right| + K_2}$$

## Boundary values

The boundary values for these equations are two:

1. For $y' = 0$, it is verify that, in *Dune II*, for the *y* variable, we have two solutions which difference is equal to $2c$. For this:

$$K_1 = \frac{R\sqrt{m_n} \cdot \sqrt[3]{4R \cdot m_{Saturn} \cdot \sqrt{k_1}}}{c} - m_{Saturn} \cdot c \cdot \sqrt{k_1} \approx -m_{Saturn} \cdot c \cdot \sqrt{k_1}$$

That is to say, *Dune II* is

$$DuneII : x^2 + y^2 = \frac{4R^4 \cdot m_n^2}{\left(m_{Saturn} \cdot (x - c) \cdot \sqrt{k_1}\right)^2}$$

It is remarkable to observe that, for $x = c$ this function is not defined and if $x > c$ the value of the radius of curvature to the square is negative. Evidently this fact must to that the profile *Dune II* is very similar to a sphere of radius $c$.

2. For $y = 0$ it verifies that, in *Dune III*, for $x$ variable there are two solutions whose difference is equal to $2a$

This solution is near $x = a$. Remembering which $k_2 \cdot a$ is the mass of the dune and its value is in the same order of magnitude than $m_n$, we can approximate:

$$\ln|m_n - k_2 x| \approx \ln(m_n) - \frac{x}{a}$$

replacing the condition 2 and this approximation in *Dune III*, we have:

$$x^2 = \frac{k_2^2 \cdot R^4}{m_{Saturn}^2 \cdot \left(\ln m_n - \frac{x}{a}\right)^2} + K_2$$

that it leads to the quadratic equation:

$$P(x) = x^4 - [2a \cdot \ln m_n] x^3 + [a^2 \cdot \ln^2 m_n] x^2 + \frac{a^2 (K_2 - k_2 R^4)}{m_{Saturn}^2}$$

For $P(x) = 0$.

The maximum, minimum and inflection point of $P(x)$ are in:

- Minimum: $x = 0$
- Inflection point: $x \approx \frac{1}{5} a \cdot \ln m_n$
- Maximum: $x = \frac{1}{2} a \cdot \ln m_n$

- Inflection point: $x \approx \dfrac{4}{5} a \cdot \ln m_n$

- Minimum: $x = a \cdot \ln m_n$

As for all these values, $P(x)$ has the same sign, the only option to have two real roots is that the independent term of $P(x)$ is negative. By numerical calculation we can obtained the two roots of $P(x)$ such that a distance $2 \cdot a$ in function of $K_2$. This value is:

$$K_2 \approx -1{,}39 \cdot 10^{65}\ kg \cdot m^3.$$

For this value, the roots are: $x_1 \approx -18.000\ m$ and $x_2 \approx 19.000\ m$ (see graph 2).

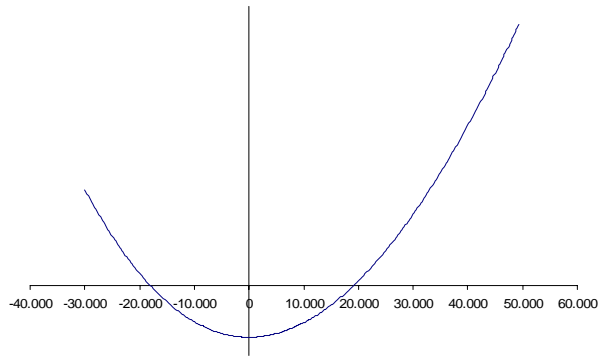

Graph 2. $P(x)$ for $K_2 \approx -1{,}39 \cdot 10^{65}\ kg \cdot m^3$.

## Analysis of the equations family

The equations family that we have obtained has the form: $(x^2 + y^2)(A \cdot x + B)^2 = K^2$

In order to analyze these equations we factorize it in the next form:

$$Dune: (x^2 + y^2) \cdot f(x) = K^2$$
$$f(x) = (A \cdot x + B)^2$$

That is to say, if $f(x)$ is constant, the *Dune* equation is reduced to a circumference. But, as $f(x)$ is a parabola, when we increases the value of the $A$ parameter, we deformed the circumference. The relation between the circumference of radius K and parabola f(x) will determine the form and gravitational stability of *Dune*.

We are going to represent a case simple to understand the physical meaning of this relation. For it, we will comment the graphs for simple values of *K*, *A* and *B*. Concretely we will take the values: *K* = 6 and *A* = 1 and we will vary the values of *B*.

a) When the value of parameter *B* is greater than *K*, parabola *f(x)* is located outside the circumference. Since the gravitational stability zone is within the circumference of radius *K*, equation *Dune* comes near to a circumference (see graph 3)

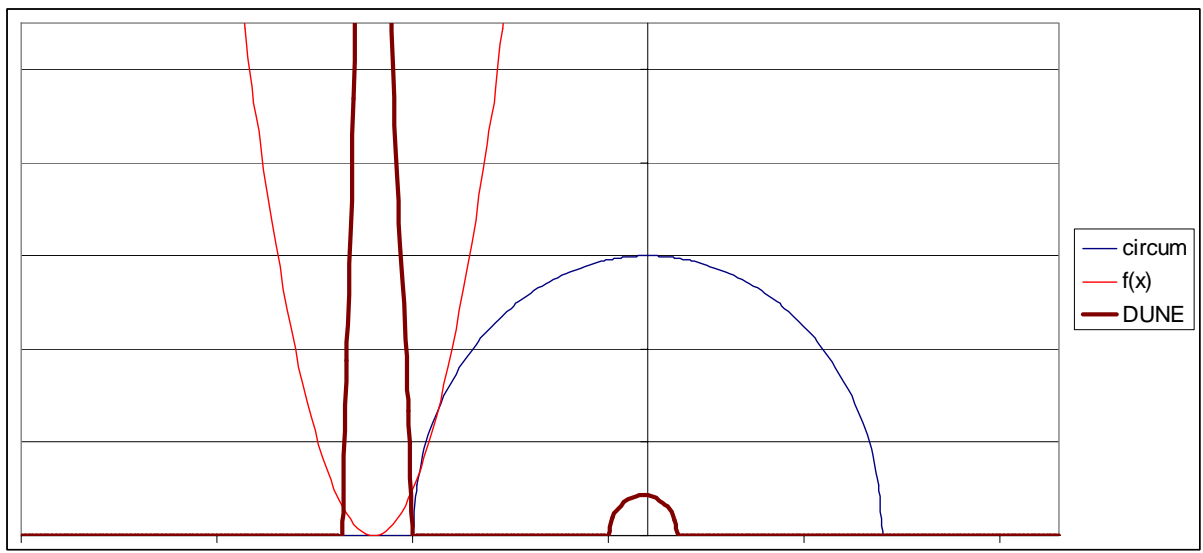

*Graph 3. Parameters K = 6, A = 1 y B = 7.*

We must observe that where parabola *f(x)* has small values it produces an asymptotic point in the *Dune* equation.

b) If parameter *B* is minor than *K* the minimum of the parabola is located within the circle of radius *K*, deforming the gravitational stability zone and transforming it into an ovoid (see graph 4).

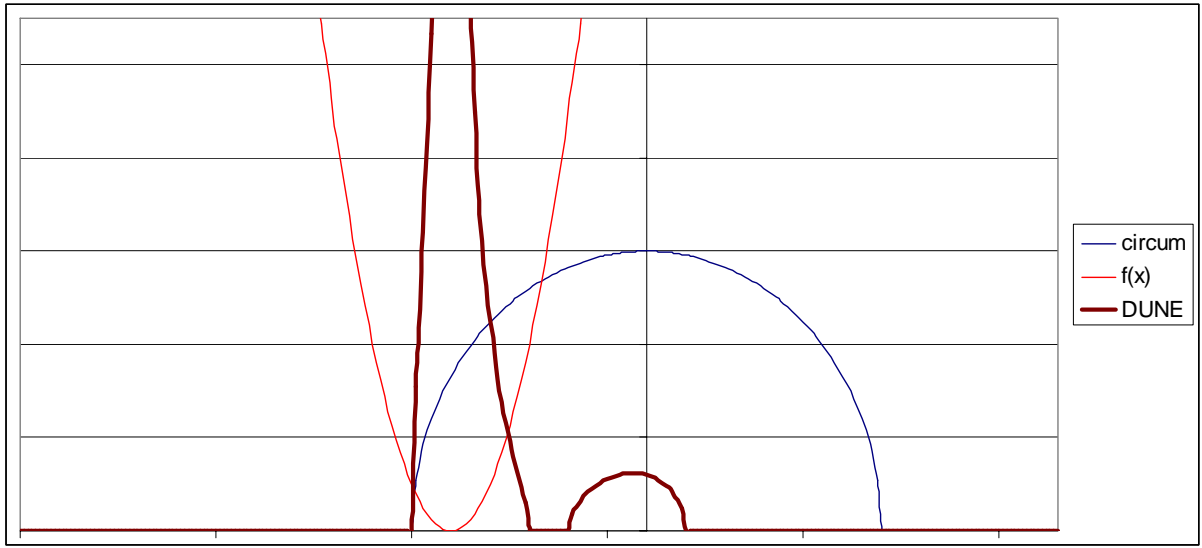

*Graph 4. Parameters K = 6, A = 1 y B = 5.*

We can observe that the asymptotic part of equation *Dune* comes near to the gravitational stability zone.

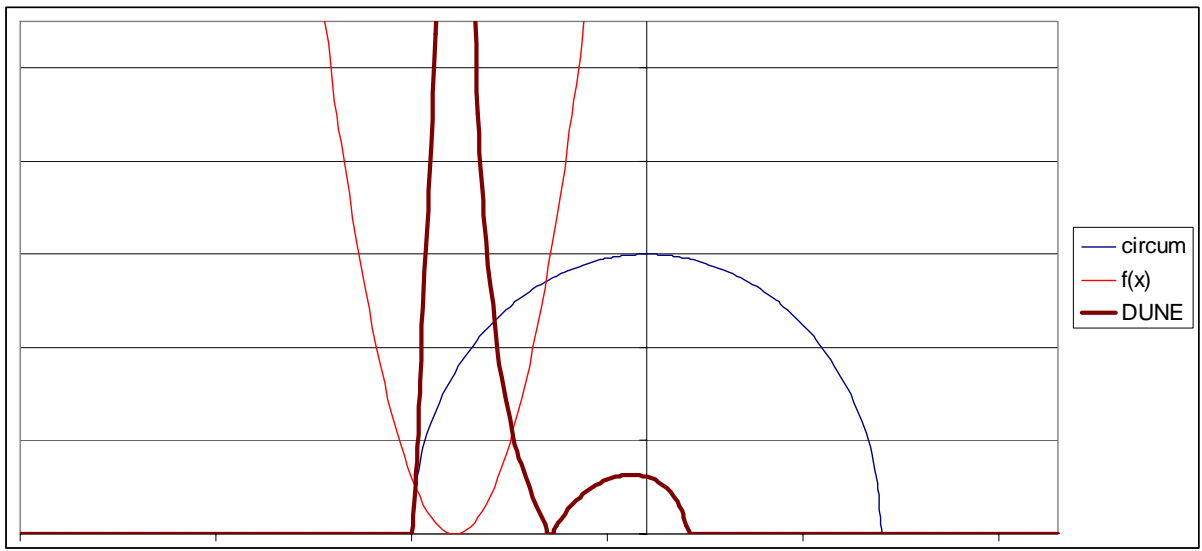

*Graph 5. Parameters K = 6, A = 1 y B = 4.9.*

c) For a concrete value, that in our case is $B = 4.9$ the gravitational stability zone and the asymptotic zone is touched in a point that coincide with the intersection of the circumference of radius *K* and parabola *f(x)* (see graph 5).

d) Once exceed the critical value of *B*, the gravitational stability zone and the asymptotic one are overlapped so that the matter of the satellite begins to lose, or more general, it can have interchange of matter between asymptotic and gravitational stability zones. This mechanism coincides with the Roche limit (see graph 6)

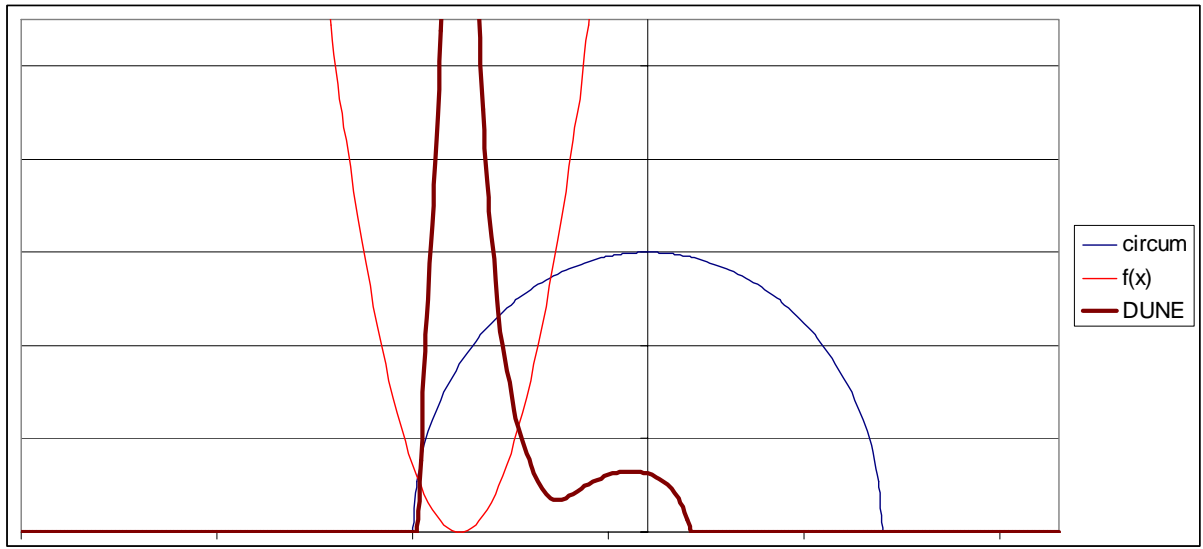

*Graph 6. Parameters K = 6, A = 1 y B = 4,8.*

This parameters relation indicates us that exists a near relation between a satellite formed by dust and a ring exists that forms from the material interchange with the satellite due to the effects of its proximity to the Roche limit.

In the particular case of the Atlas satellite its closey linked with the small ring R/2004 S 1 can be explained by this mechanism.

## Atlas graphical representation.

The equations that we have obtained for the "flying dune" Atlas model are the following ones:

$$Rock-nucleus: \frac{x^2}{r_{rock}^2} + \frac{y^2}{c^2} = 1$$

$$DuneII: (x^2+y^2)\left(m_{Saturn} \cdot (x-c) \cdot \sqrt{k_1}\right)^2 = 4R^4 \cdot m_n^2$$

$$DuneIII: (x^2+y^2)\left(m_{Saturn}^2 \cdot \ln^2|m_n - k_2 \cdot x| + K_2\right) = k_2^2 \cdot R^4$$

The parameters $r_{rock}$, $m_n$, $k_1$, $k_2$ and $K_2$ we know them only by suppositions. Nevertheless, an adequate adjustment of these parameters leads to a graphical representation of the following type:

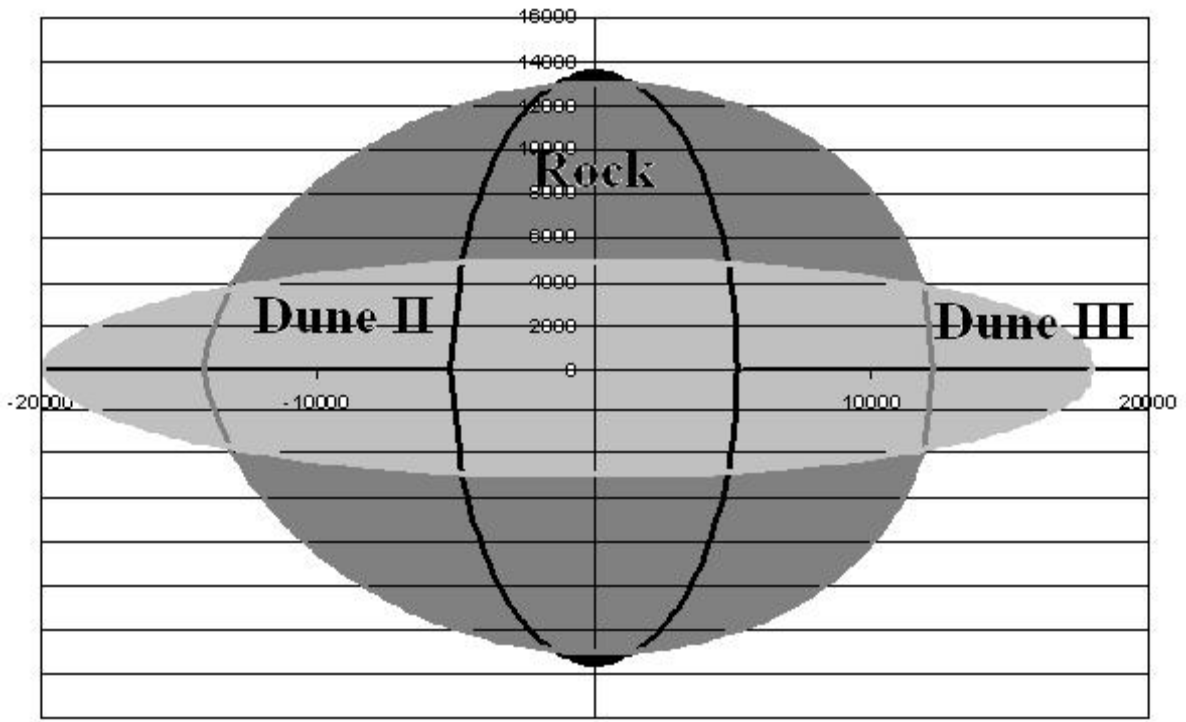

*Figure 4. Graphical Atlas representation according to the "flying dune" model.*

That we can compare with the images taken by the Cassini prove:

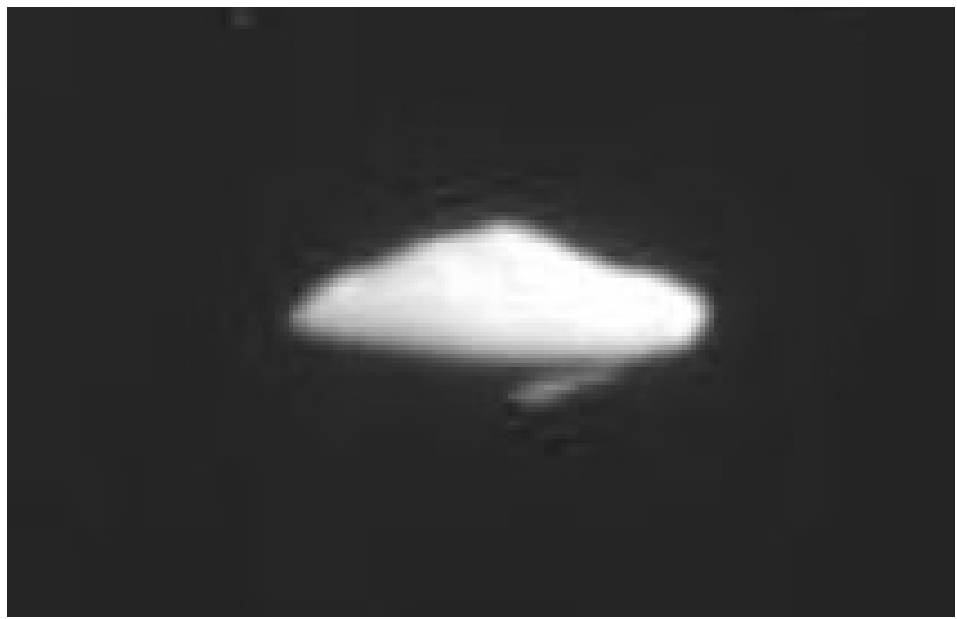

*Imagen 8. Atlas seen in an enlargement of the photography from June 8, 2005. (Credit: NASA, JPL, SSI)*

# Conclusions.

- The analysis of the Roche limit demonstrate that it is not possible that Atlas were a totally deformable satellite, but the images that show a symmetry according to the polar axis indicate that it is not totally rigid.

- The atypical Atlas form, can be explained by a model that a rock nucleus combines with a dust dune that moves in its equatorial zone.

- An exact determination of their mass, composition and other parameters would permit us obtain an exact representation of their form by means of computer science models. By our estimations Atlas mass is slightly superior to $2.76 \cdot 10^{+15}$ *kg* and it composition approximately is $m_{rigid}$ ~ 43,15 % y $m_{deformable}$ ~ 56,85 %, such a way that, it is located nearly in the Roche limit.

- The Atlas rotation (synchronous) and the tidal forces of Saturn are displaced the dust cloud towards the equator forming a dense dune and undressing the poles, in which it is possible to be seen the irregularities of the rock nucleus.

- Since Atlas sees perturbed by the gravitational fields from other objects of the Saturn system, the dune must have a slight undulatory movement in its surface.

- We can observe a mathematical relation between a satellite that contains a dust "dune" and the presence of a small ring with which to interchange matter, as it can be the case of Atlas and the small ring R/2004 S 1.

- If Cassini probe, passed a time period, returned to take images from Atlas from a polar plane, with same resolution as June 12, 2007, we could compare with first for verifying if significant differences exist.

# Acknowledgement

The model presented here arose as result of the different opinions expressed in an astronomy forum on the web page: www.sondasespaciales.com.

Firstly, I want to thank to Pedro Leon (nick: Pedro) because he have created this marvelous place of exchange information and ideas about Astronomy.

Secondly, I wanted to thank for his aid to Adolph Reig García-San Pedro (nick: Adonis). He made the first rough draft about the analysis of the forces on the Atlas surface. Some parts of my analysis are based on qualify and completing this rough draft. As well as Newton said in an occasion: "*If I have seen a little further it is by standing on the shoulders of Giants*."

The Internet forums are an excellent way to implement the technique of "brainstorming" that so good results offers in the management companies to solve complex problems. For that reason I want to thank to each and every one of which they were there, in the thread **Atlas, sin duda un nuevo misterio,** first its ideas and later its aid, support and spirit so that this model got to take shape in the present article.

Thanks to: Manuel Marqués López (nick: nimbar), David Mayo Turrado (nick: Mayo), Luis Gascón (nick: Toulouse), Pedro León (nick: Pedro), Adolfo Reig García-San Pedro (nick: Adonis), Javier Baena (nick: urheimait), nick: neotrantoriano (who first put to us on the track of the "*flying dune*"), David Vilches (nick: tucker), nick: Manu, Aitor Conde (nick: Bultza) who was first in animating to me to write a report on the ideas that were being considering about Atlas, José Andrés Pérez (nick: zeroauriga), Augusto Bibolotti (nik: Eliolari), José María Ruiz Moreno (nick: Spiri), José Sánchez (nick: Rick Dekkard) and Sergio Álvarez Sanchís (nick: sergiotas).

Thanks to everybody.

# References.

# Bibliography.